\documentstyle[preprint,aps]{revtex}

\tightenlines

\begin{document}
\def\W{{\cal W}}
\def\U{{\cal U}}
\def\T{{\cal T}}
\def\R{{\cal R}}
\def\D{{\Delta}}
\def\tD{{\tilde\Delta}}
\newcommand{\ket}[1]{|\,#1\,\rangle}
\newcommand{\bra}[1]{\langle\,#1\,|}
\newcommand{\braket}[2]{\langle\,#1\,|\,#2\,\rangle}

\title{Wigner Functions on a Lattice}
\author{Akiyoshi Takami, Takaaki Hashimoto, Minoru Horibe 
            and Akihisa Hayashi}
\address{Department of Applied Physics\\
           Fukui University, Fukui 910, Japan}
\draft

\maketitle

\begin{abstract}
  The Wigner functions on the one dimensional lattice are studied.
Contrary to the previous claim in literature, 
Wigner functions exist on the lattice 
with any number of sites,
whether it is even or odd.
There are infinitely many solutions satisfying the conditions 
which reasonable Wigner functions should respect.
After presenting a heuristic method to obtain Wigner functions,
we give the general form of the solutions. Quantum mechanical
expectation values in terms of Wigner functions are 
also discussed. 
\end{abstract}
\pacs{PACS: 01.55.+b; 03.65.-w; 03.65.Sq}

\section{Introduction}
The Wigner function, the Wigner transform of the density matrix, 
has been recognized as a bridge between quantum
and classical mechanics \cite{wigner}.  
It is the quantum mechanical counterpart to the 
distribution function in the classical phase space:  
the Wigner function $\W(q,p)$ gives the position distribution
if integrated over the momentum $p$ and the momentum distribution
if integrated over the coordinate $q$. Though the Wigner function
may be negative, the equations of motion for 
$\W(q,p)$, in the classical limit, are reduced to the classical
Liouville equation for the distribution function.
The Wigner functions have been used to analyze 
various phenomena where the quantum and classical correspondence  
is crucial \cite{ring}\cite{friesch}. One of the important and
interesting applications of the Wigner functions is the 
quantum tomography, where the Wigner function $\W(q,p)$ for 
a certain quantum system is measured for any $q$ and $p$ and thus 
the full quantum information of the system,
density matrix, can be reproduced \cite{vogel}\cite{smithey}. 

Some time ago Wootters \cite{wootters} and
Cohendet et. al. \cite{cohendet} studied Wigner functions for 
a finite discrete space like a spin space. 
For an integer spin $s$ in a pure state,
they found a possible Wigner function is given by 
\begin{eqnarray}
\W(a,b) = \frac{1}{2s+1}
            \sum_{h=-s}^{s} \exp\left( i\frac{4\pi}{2s+1}hb \right)
            \psi^*(a+h) \psi(a-h),\ (a,b=-s, -s+1,\ldots , s), 
          \label{eq_cohendet1}
\end{eqnarray}
where $\psi(\alpha)$ is the amplitude for each 
spin state labeled by an integer $\alpha=-s, -s+1,\ldots,s$ and 
$\psi(\alpha + 2s+1)=\psi(\alpha)$ is assumed.  
Their results, however, can not be applied to half-integer
spin systems. 
The total number of sites is $N\equiv 2s+1$ and 
the Wigner function given in
\cite{wootters} and \cite{cohendet} 
is only for the lattice with odd number sites. 

Leonhardt proposed the Wigner function for an even number of sites
$N$ and discussed applications to the quantum tomography in a 
finite discrete space \cite{leonhardt}.
However, the Wigner function given in \cite{leonhardt} involves
"ghost variables" with vanishing probability.  
Thus the construction of Wigner functions on a discrete
space or lattice has still been an open problem. 
The relations between various quasidistribution functions including
Wigner functions  and the finite discrete space 
with arbitrary Floquet angles have also been discussed 
\cite{opatrny}\cite{rivas}.

In this paper, we relabel the spin state $\alpha$ by an
integer value $a=s+\alpha$ with $a$ ranging from $0$ to $2s\equiv N-1$
 (the total number of sites is $N\equiv 2s+1$).
The equivalence between these two labeling will
be discussed in Sec.~5 in detail. 
Labeling each site by an integer $(0,1,\ldots , N{-}1)$,  
the Wigner function of Wootters and Cohendet et. al. is given by
\begin{eqnarray}
 \W(a,b) = \frac{1}{N}
            \sum_{h=0}^{N-1} \exp\left( i\frac{4\pi}{N}hb \right)
            \phi^*(a+h) \phi(a-h)         
            ,\ (a,b = 0,1,\cdots,N-1), 
            \label{eq_cohendet2}
\end{eqnarray}
where $\phi(a+N)=\phi(a)$ is assumed.
Generally Wigner functions are written in terms of the Fano
operators (matrices) $\Delta(a,b)$ \cite{fano} as 
\begin{eqnarray}
  \W(a,b) = \frac{1}{N}\sum_{a_1 a_2=0}^{N-1}   
              \phi^*(a_1) \Delta_{a_1 a_2}(a,b) \phi(a_2)
            ,\ (a,b = 0,1,\cdots,N-1), 
       \label{eq_fano}
\end{eqnarray}
or in terms of the density matrix $\rho$ as
\begin{eqnarray}
  \W(a,b) = \frac{1}{N}{\rm Tr} \rho \Delta(a,b).
         \label{eq_den1}
\end{eqnarray}

In this paper we postulate the following four 
conditions (A)$\sim$(D) for eligible Wigner functions and corresponding
Fano operators.
\begin{itemize}
\item[(A)]  The Wigner functions reproduce the distribution
            in the configuration space:
\begin{eqnarray}
   \sum_{b=0}^{N-1} \W(a,b) &=& | \phi(a)|^2,
\end{eqnarray}
or equivalently
\begin{eqnarray}
   \sum_{b=0}^{N-1} \Delta_{a_1 a_2}(a,b) &=& 
       N\delta_{a,a_1}\delta_{a_1,a_2}.
\end{eqnarray}
\item[(B)] The Wigner functions reproduce the distribution
             in the momentum space:
\begin{eqnarray}
   \sum_{a=0}^{N-1} \W(a,b) = | \tilde\phi(b) |^2,
\end{eqnarray}
where the Fourier transform $\tilde\phi$ is given by
\begin{eqnarray}
 \tilde\phi(b) = \frac{1}{\sqrt{N}}
                  \sum_{a=0}^{N-1} \exp\left(-i\frac{2\pi}{N}ba \right) 
                  \phi(a),\ (b=0,1,\ldots,N-1),
\end{eqnarray}
or equivalently
\begin{eqnarray}
  \sum_{a=0}^{N-1} \Delta_{a_1 a_2}(a,b) &=&
        \exp\left(i\frac{2\pi}{N}b(a_1-a_2)\right),
\end{eqnarray}
\item[(C)] The Wigner functions are real. Therefore the
             Fano operators are hermitian:
\begin{eqnarray}
    \W(a,b)^\ast = \W(a,b),
     \ {\rm or}\ 
     \Delta(a,b)^+ = \Delta(a,b).
\end{eqnarray}
\item[(D)]  The Fano operators are orthonormalized and 
             complete: 
\begin{eqnarray}
 {\rm Tr}\left( \Delta^+(a,b)\Delta(a^\prime,b^\prime) \right)
     &=& N\delta_{aa^\prime}\delta_{bb^\prime}.
\end{eqnarray}
\end{itemize}  
The completeness in the last condition (D) is necessary for 
that the density matrix can be expressed in terms of the 
Wigner functions. The completeness, together with the
orthonormality, enables us to obtain the simple form:
\begin{eqnarray}
   \rho_{a_1,a_2} = \sum_{a,b=0}^{N-1} \W(a,b)\Delta_{a_1 a_2}(a,b).
         \label{eq_den2}
\end{eqnarray}
We note that Eqs.(\ref{eq_den1}) and (\ref{eq_den2}) are valid
for a mixed state as well.
It is easy to see that these four conditions (A)$\sim$(D) are satisfied
by the Wigner function given by Wootters and Cohendet et. al. 

In this paper we will show that the Wigner functions with 
the properties (A)$\sim$(D) exist for the lattice with any number 
of sites, and therefore any spin system. 
We do not need any "ghost variables" as in \cite{leonhardt}.
In fact there are
infinitely many solutions and we will give the general form. 
In Sec.~2 we will first give a heuristic way to obtain 
a solution for even site systems. The general form 
for an arbitrary number of sites will be given in Sec.~3. 
In Sec.~4 we will discuss expectation values of physical
quantities in terms of Wigner functions and 
the problem on the two ways of site labeling.

\section{Construction of a Wigner function on an even site lattice}
In this section, we describe a heuristic method to construct a
Winger function
on an even site lattice and show its existence.
Let \(N\) be an even integer and \({\bf Z}_N\) be the cyclic group
whose elements are labeled from \(0\) to \(N-1\).
We construct the \(N\times N\) Fano operators by a reduction method
starting from \(2N\times2N\) matrices.
The enlarged matrices have a similar form to a cyclic representation
of the Weyl matrices on an odd site lattice.

We define phase and shift operators $P$ and $S$ as
\begin{equation}
P_{a_1 a_2}=\omega'^{a_1}\delta_{a_1 a_2},
\hspace{8mm}
S_{a_1 a_2}=\delta_{a_1,a_2-1},
\end{equation}
and two operators \(T\) and \(B\) as
\begin{equation}
T_{a_1 a_2}=\delta_{a_1,2N-1-a_2},
\hspace{8mm}
B_{a_1 a_2}=\delta_{a_1,a_2+(-1)^{a_2}},
\end{equation}
which are \(2N\times2N\) matrices,
where \(a_1,a_2\in {\bf Z}_{2N}\) and \(\omega'\) is the primitive
\(2N\)-th root of unity:
\begin{equation}
\omega'=\exp\left({{2\pi i}\over{2N}}\right).
\end{equation}
These operators have the following properties:
\begin{eqnarray}
& & P^+=P^{-1}=P^*,\ S^+=S^{-1}=S^t,\ T^+=T^{-1}=T=T^*,
            \nonumber \\
& & PS=\omega'^{-1}SP,\ TS=S^{-1}T,\ BT=TB.
\end{eqnarray}
We consider a set of matrices \(\overline{W}(a,b)\) constructed
from these matrices:
\begin{eqnarray}
\overline{W}(a,b) &=& \omega'^{-ab}P^bS^{-(2a+1)},
             \nonumber \\
\overline{W}(a+N,b) &=& \omega'^{-(a+N)b}P^bS^{-(2a+1)}B,
\end{eqnarray}
where \(a\in{\bf Z}_{N}, b\in{\bf Z}_{2N}\).
They are orthogonal in the trace norm and complete in the
\(2N\times2N\)-dimensional operator
space, but not hermitian.
We define matrices \(\overline{\Delta}(a,b)\) by
\begin{equation}
\overline{\Delta}(a,b)=\overline{W}(a,b)T,
\end{equation}
where \(a, b\in{\bf Z}_{2N}\).
For these operators $\overline\Delta(a,b)$, it can be shown that
the condition (D) in the previous section is fulfilled and
the following properties (A'), (B') hold.

\noindent
(A') A property which is similar to (A) in the previous section:
\begin{eqnarray}
\sum_{b=0}^{2N-1}\overline{\Delta}_{a_1 a_2}(a,b) &=&
          {N\delta_{aa_1}\delta_{a_1a_2}},
                     \nonumber \\
\sum_{b=0}^{2N-1}\overline{\Delta}_{a_1 a_2}(a+N,b) &=&
       {N\delta_{aa_1}\delta_{a_1a_2}},
\end{eqnarray}
\noindent
\ \ \ \ \ \ \ where \(a\in{\bf Z}_{N}\). \\

\noindent
(B') A property which is similar to (B) in the previous section:
\begin{eqnarray}
\sum_{a=0}^{2N-1}\overline{\Delta}_{a_1 a_2}(a,b) &=& 
    \exp\left(i{2\pi\over
    N}b(a_1-a_2)\right), \hspace{5mm} {\rm for\ even}\ b
               \nonumber \\
\sum_{a=0}^{2N-1}\overline{\Delta}_{a_1 a_2}(a,b) &=& 0,
       \hspace{10mm} {\rm for\ odd}\ b
\end{eqnarray}
\ \ \ \ \ \ \ where \(b\in{\bf Z}_{2N}\). \\

\noindent
Note that the Weyl matrices on an odd lattice, i.e. for odd
\(N\),  can be represented as
\begin{equation}
W(a,b)=\omega^{-2ab}P^{2b}S^{-2a}=S^{-a}P^{2b}S^{-a},
\end{equation}
using similarly defined \(N\times N\) phase and shift operators,
where \(\omega\) is the primitive \(N\)-th root of unity.

We restrict the complex \(2N\)-dimensional vector space,
on which the \(\overline{\Delta}(a,b)\) matrices act,
to the \(N\)-dimensional \(B\)-invariant subspace \(V\):
\begin{equation}
V=\{\phi\in{\bf C}^{2N}: B\phi=\phi \},
\end{equation}
which means \(\phi(2a)=\phi(2a+1), a\in{\bf Z}_N\)
if \(\phi\in V\).
The matrices \(\overline{\Delta}(a,b)\) take \(N\times N\)-dimensional
form on this restricted space,
e.g. choosing \(\phi(2a), a\in{\bf Z}_N\) as independent ones,
and we denote the reduced matrices by \(\Delta'(a,b)\).
The hermiticity is recovered after the reduction.
It can be seen that the next two identities hold for the \(\Delta'\),
\begin{equation}
\Delta'(a,b)=(-1)^b\Delta'(a+N,b),
\end{equation}
\begin{equation}
\langle\Delta'(2a,b),a\in{\bf Z}_N,b\in{\bf Z}_{2N}\rangle=
\langle\Delta'(2a+1,b),a\in{\bf Z}_N,b\in{\bf Z}_{2N}\rangle,
\end{equation}
where \(\langle\ \rangle\) means the vector space spanned by
the elements between the bra and ket.
Only \(N^2\) matrices out of \(4N^2\) matrices of \(\Delta'\) are
independent.
We define Fano operators as
\begin{equation}
\Delta(a,b)={1\over2}\left\{\Delta'(2a,2b)+\Delta'(2a,2b+1)\right\},
\end{equation}
where \(a, b\in{\bf Z}_{N}\).
We can see that these operators satisfy the properties (A)\(\sim\)(D).
The explicit form of the corresponding Wigner functions is given by
\begin{eqnarray}
\W(a,b)&=&{1\over2N}\sum_{h=0}^{N-1}
(\omega'^{4hb}+\omega'^{2h(2b+1)})\phi ^*(a+h)\varphi(a-h)
           \nonumber \\
&+&{1\over2N}\sum_{h=0}^{N-1}(\omega'^{(2h+1)2b}+
\omega'^{(2h+1)(2b+1)})\phi^*(a+h)\phi(a-h-1). \label{eq_evensol}
\end{eqnarray}

\section{General form of Wigner functions on a lattice}
In this section, we will give the general form for 
Wigner functions on the lattice with arbitrary lattice size $N$.
As in the previous sections, we assume that the Wigner function 
should satisfy the four conditions (A)$\sim$(D) given in Sec.~1.  

It turns out that the Fourier transforms of the Fano operators 
are much convenient.
\begin{eqnarray}
    \tD(n,m) &=& \frac{1}{N} \sum_{a,b=0}^{N-1}
             e^{i\frac{2\pi}{N}\left( na + mb \right)}
             \D(a,b),\ (n,m=0,1,\ldots,N-1),   \nonumber \\
    \D(a,b) &=& \frac{1}{N} \sum_{n,m=0}^{N-1}
             e^{-i\frac{2\pi}{N}\left( na + mb \right)}
             \tD(n,m),
       \label{eq_tDfourie}
\end{eqnarray}
where the site suffixes are suppressed.
In terms of these Fourier transforms $\tD$, the four conditions 
take simple forms: 
\begin{eqnarray}
    \tD_{a_1 a_2}(n,0) &=& 
          e^{i\frac{2\pi}{N}n a_1}\delta_{a_1,a_2},
                    \nonumber \\
    \tD_{a_1 a_2}(0,m) &=& \delta_{-a_1+a_2,m},
                    \nonumber \\
    \tD^+(n,m) &=& \tD(N-n,N-m),
                    \nonumber \\
    {\rm Tr}\left( \tD(n,m)^+
                   \tD(n^\prime,m^\prime) \right)
     &=& N\delta_{n,n^\prime}
                    \delta_{m,m^\prime}. \label{eq_tD}
\end{eqnarray}

We will see there exist infinitely many solutions to the 
above non-linear equations Eqs.(\ref{eq_tD}). 
We first show that general solutions can be written
in terms of any special solution and an 
arbitrary $(N-1)^2\times(N-1)^2$ orthogonal matrix. Construction
for special solutions will be given later.

>From Eqs.(\ref{eq_tD}) we first notice that $\tD(n,0)$
and $\tD(0,m)$ are unique. 
Denoting a special solution by $\tD_0$ and an arbitrary solution 
by $\tD$, we have 
\begin{eqnarray}
   \tD_{a_1 a_2}(n,0) &=& \tD_{0\,a_1 a_2}(n,0) = 
          e^{i\frac{2\pi}{N}n a_1}\delta_{a_1,a_2},
                      \nonumber \\
   \tD_{a_1 a_2}(0,m) &=& \tD_{0\,a_1 a_2}(0,m) = 
          \delta_{-a_1+a_2,m}.     \label{eq_tD0}
\end{eqnarray}
These relations and the orthonormal completeness  
of $\tD$ imply, 
for the remaining suffixes $n,m,n^\prime,m^\prime = 1,2,\ldots,N-1$,
each $\tD(n,m)$ is expressed by a linear combination of  
$\tD_0(n^\prime,m^\prime)$'s. Thus we have
\begin{eqnarray}
 \tD(n,m) = \sum_{n^\prime,m^\prime=1}^{N-1}
    \U_{nm;n^\prime m^\prime}
             \tD_0(n^\prime,m^\prime),
       \ \ (n,m,n^\prime,m^\prime = 1,2,\ldots,N-1). 
            \label{eq_tDlinear}
\end{eqnarray}
It can be easily shown that $\tD$ is also a solution to 
Eqs.(\ref{eq_tD})
, if and only if  the complex $(N-1)^2\times(N-1)^2$ matrix 
$\U_{nm;n^\prime m^\prime}$
satisfies the conditions:
\begin{eqnarray}
    \U^\ast_{nm;n^\prime m^\prime} 
    &=& \U_{N-n,N-m;N-n^\prime,N-m^\prime},
                \nonumber \\
    \U^+ &=& \U^{-1}.   \label{eq_Ucond}
\end{eqnarray}
These unfamiliar conditions Eqs.(\ref{eq_Ucond}) for the matrix
$\U$ can be expressed in a more familiar form by use of  
an $(N-1)^2\times(N-1)^2$ orthogonal matrix. We introduce 
an $(N-1)^2\times(N-1)^2$ matrix $\T$ as
\begin{eqnarray}
   \T_{nm;n^\prime m^\prime}
        = \delta_{n,N-n^\prime}
          \delta_{m,N-m^\prime},
     \ \ (n,m,n^\prime,m^\prime = 1,2,\ldots,N-1)
\end{eqnarray}
with the following properties:
\begin{eqnarray}
    \T = \T^\ast = \T^+ = \T^{-1}.
\end{eqnarray}
The conditions Eqs.(\ref{eq_Ucond}) take compact forms as
\begin{eqnarray}
     \U^\ast = \T\U\T,\ \U^+ = \U^{-1}.   \label{eq_Ucond_new}
\end{eqnarray}
And using the unitary matrix $(1+i\T)/\sqrt{2}$,
we write
\begin{eqnarray}
    \U = \frac{1+i\T}{\sqrt{2}}\R\frac{1-i\T}{\sqrt{2}},
\end{eqnarray}
where $\R$ is an $(N-1)^2\times(N-1)^2$ matrix.
Now it is not difficult to show that the matrix $\U$ satisfies the
conditions Eqs.(\ref{eq_Ucond_new}), if and only if the 
matrix $\R$ is real and orthogonal. 

In conclusion, 
as for the Fourier transforms of the Fano operators with the 
four properties (A)$\sim$(D), 
$\tD(n,0)$ and $\tD(0,m)$ are uniquely given as 
in Eqs.(\ref{eq_tD0}) and 
the remaining other ones are given  in terms of a special
solution $\tD_0$ and an $(N-1)^2\times(N-1)^2$
real orthogonal matrix $\R$ as follows:
\begin{eqnarray}
 \tD(n,m) &=&\sum_{n^\prime,m^\prime=1}^{N-1}
    \U_{nm;n^\prime m^\prime}
             \tD_0(n^\prime,m^\prime),
    \ \ (n,m,n^\prime,m^\prime = 1,2,\ldots,N-1), 
                     \nonumber \\
    \U &=& \frac{1+i\T}{\sqrt{2}}\R\frac{1-i\T}{\sqrt{2}},
\end{eqnarray}
where 
\begin{eqnarray}
    \R^{{\rm T}} &=& \R^{-1},
                     \nonumber \\
   \T_{nm;n^\prime m^\prime}
        &=& \delta_{n,N-n^\prime}
          \delta_{m,N-m^\prime},
    \ \ (n,m,n^\prime,m^\prime = 1,2,\ldots,N-1). 
\end{eqnarray}

Now we will construct special solutions to Eqs.(\ref{eq_tD}) and 
discuss how they are related to solutions given before: 
the solution for odd $N$ given by Cohendet et. al. and 
the solution for even $N$ obtained by the reduction method in this
paper.

As we have shown in Eqs.(\ref{eq_tD}), $\tD(n,0)$
and $\tD(0,m)$ are unique. They are most conveniently expressed as 
\begin{eqnarray}
    \tD(n,0) = P^n, \nonumber \\
    \tD(0,m) = S^m,
\end{eqnarray}
where $P$ is the phase operator and $S$ is the shift
operator, which were introduced in the previous section:   
\begin{eqnarray}
    P_{a_1 a_2}&=& e^{i\frac{2\pi}{N}a_1}\delta_{a_1,a_2},
                      \nonumber \\
    S_{a_1 a_2}&=& \delta_{a_1,a_2-1}. 
          \label{eq_tD0ps}
\end{eqnarray}
We can show that the set of operators 
$P^n S^m (n,m = 0,1,
\ldots,N-1)$ forms a orthonormal complete set as
\begin{eqnarray}
    {\rm Tr}(P^n S^m)^+
            (P^{n^\prime} S^{m^\prime} )
    = N\delta_{n^\prime n}\delta_{m^\prime m}.
          \label{eq_complete}
\end{eqnarray}
with the property under hermitian conjugation: 
\begin{eqnarray}
    \left( P^n S^m \right)^+ = 
       e^{i\frac{2\pi}{N}nm}P^{N-n}S^{N-m}.
            \label{eq_hermite}
\end{eqnarray}
The relations Eqs.(\ref{eq_tD0ps}), (\ref{eq_complete}) 
and (\ref{eq_hermite}) imply that a special solution 
may be given by the form:
\begin{eqnarray}
   \tD_0(n,m) = e^{i\theta(n,m)}P^n S^m,
        \label{eq_tD0form}
\end{eqnarray}
if the phase $\theta(n,m)$ can be properly chosen
so that the condition $\tD^+(n,m) = \tD(N-n,N-m) $ holds. 
The condition for the phase $\theta$
turns out to be
\begin{eqnarray}
     \theta(n,m) \equiv -\theta(N-n,N-m)
                            +\frac{2\pi}{N}nm,
                               \ \ ({\rm mod}\ 2\pi). 
\end{eqnarray}

We will give some examples.  The solution for odd $N$ given
by Cohendet et. al.  is reproduced by the choice: 
\begin{eqnarray} 
   \theta(n,m) = 
   \cases{ \frac{\pi}{N}nm & ($m=$\,even) \cr 
           \frac{\pi}{N}(N-n)(N-m) & ($m=$\,odd) \cr}.
\end{eqnarray} 
After the Fourier inverse transformation, this gives the 
Fano operator for odd $N$ as
\begin{eqnarray}
 \Delta_{a_1 a_2}(a,b) = e^{i\frac{2\pi}{N}b(a_1-a_2)}
           \delta_{2a,a_1+a_2},
\end{eqnarray}
and the Wigner function given in Eq.(\ref{eq_cohendet2}).

The solution by the reduction method for even $N$ in Sec.~2 corresponds
to the choice: 
\begin{eqnarray} 
   \theta(n,m) = 
   \cases{ \frac{\pi}{N}nm & ($n=$\,even,\ $m=$\,even) \cr 
           \frac{\pi}{N}(nm+n) & ($n=$\,even,\ $m=$\,odd ) \cr
           \frac{\pi}{N}(nm-m) & ($n=$\,odd ,\ $m=$\,even) \cr
           \frac{\pi}{N}(nm+n-m) & ($n=$\,odd ,\ $m=$\,odd ) \cr},
\end{eqnarray} 
which gives the Wigner function given by Eq.(\ref{eq_evensol}).

\section{Discussions} 
We have postulated the four conditions (A)$\sim$(D) on 
the Wigner functions
on the lattice. After presenting a heuristic reduction method
to obtain a special solution on an even site lattice, we found
the general form of the solutions for any size $N$
, which is parameterized by an arbitrary orthogonal matrix.

Now we will discuss how the quantum mechanical expectation value
of a physical quantity is related to the Wigner functions. 
Let us introduce the position, site, operator $\hat q$ and 
the momentum operator $\hat p$ on the lattice as follows:
\begin{eqnarray}
    \hat q_{a_1 a_2}&=&a_1\delta_{a_1,a_2}, \nonumber \\
    \hat p_{a_1 a_2}&=&\frac{1}{N}\sum_{b=0}^{N-1}
                         be^{i\frac{2\pi}{N}b(a_1-a_2)}.
\end{eqnarray}
From the conditions (A) and (B), it is evident that the 
expectation value of the physical quantity like $\hat q^n$ or
$\hat p^m$ are expressed as
\begin{eqnarray}
    {\rm Tr}\,\rho \hat q^n &=& \sum_{a,b=0}^{N-1}a^n \W(a,b), 
             \nonumber \\
    {\rm Tr}\,\rho \hat p^m &=& \sum_{a,b=0}^{N-1}b^m \W(a,b).
\end{eqnarray}
Differences in many solutions arise, when we evaluate  
the expectation value for  
a more general quantity involving both $\hat q$ and $\hat p$.
For a general physical quantity, however, it is more convenient to 
use the phase and shift operators $P$ and $S$ 
instead of $\hat q$ and $\hat p$: 
\begin{eqnarray}
    P&=&e^{i\frac{2\pi}{N}\hat q}, \nonumber \\
    S&=&e^{i\frac{2\pi}{N}\hat p},
\end{eqnarray}
whose matrix elements are given in the previous
section. In view of the finiteness of the lattice, the use of
$P$ and $S$ is more natural than $\hat q$ and $\hat p$. 
Now we calculate the average of the classical quantity $P^nS^m$ with 
the Wigner function being the weight, the distribution function, as 
\begin{eqnarray}
   \langle P^nS^m \rangle_{{\rm Wigner}} \equiv 
   \sum_{a,b=0}^{N-1} \left( e^{i\frac{2\pi}{N}a} \right)^n
                      \left( e^{i\frac{2\pi}{N}b} \right)^m
                      \W(a,b).
\end{eqnarray}
From Eq.(\ref{eq_den1}) and Eqs.(\ref{eq_tDfourie}),
we find that the right hand side of this equation 
is nothing but the expectation value of the Fourier transform
of the Fano operator, and thus we have 
\begin{eqnarray}
   \langle P^nS^m \rangle_{{\rm Wigner}} = {\rm Tr}\,\rho\tD(n,m).
\end{eqnarray}
As we have seen in the previous section, there are infinitely many
solutions for $\tD$'s. If we take the 
form as in Eq.~(\ref{eq_tD0form}):
\begin{eqnarray}
    \tD(n,m) =  e^{i\theta(n,m)}P^n  S^m, 
\end{eqnarray}
then we have
\begin{eqnarray}
   \langle P^nS^m \rangle_{{\rm Wigner}} = 
    e^{i\theta(n,m)}\,{\rm Tr}\,\rho P^n S^m.
\end{eqnarray}
With other solutions for $\tD$, we would obtain different 
operators on the right-hand side.
Detailed general discussions on this matter, including the 
continuum limit of the lattice Wigner functions, 
will be given elsewhere \cite{horibe}.  

Next we clarify the equivalence between the two ways of 
site labeling mentioned in Sec.~1. 
In the labeling adopted in this paper, each site 
is specified by an integer $a=0,1,\ldots,N-1$ with the 
amplitude $\phi(a)$.  
In another way of labeling, which is more appropriate for a 
system of spin $s$ (integer or half-integer) 
, we label each site by $\alpha =-s,-s+1,\ldots,s$ 
with the amplitude $\psi(\alpha)$. 
The total number of sites is $N=2s+1$.
The correspondence satisfying $\psi(\alpha)\sim \phi(a)$
and $ \tilde\psi(\beta)\sim \tilde\phi(b)$ is, up to a constant
phase factor, given by
\begin{eqnarray} 
  \psi(\alpha) &=& e^{-i\frac{2\pi}{N}as} \phi(a),
        \ \ \alpha = -s+a, \nonumber \\ 
  \tilde\psi(\beta) &=& e^{i\frac{2\pi}{N}\beta s} \tilde\phi(b),\ 
        \ \ \beta = -s+b, \nonumber \\
   & &(a,b=0,1,\ldots,N-1;\ \alpha,\beta=-s,-s+1,\ldots,s), 
             \label{eq_corres}
\end{eqnarray}
provided that the Fourier transforms $\tilde\psi$ and $\tilde\phi$
are most naturally defined as
\begin{eqnarray}
 \tilde\phi(b) &=& \frac{1}{\sqrt{N}}
                  \sum_{a=0}^{N-1} e^{-i\frac{2\pi}{N}ba} 
                  \phi(a),\ \ (b=0,1,\ldots,N),
                         \nonumber  \\
 \tilde\psi(\beta) &=& \frac{1}{\sqrt{2s+1}}
                  \sum_{\alpha=-s}^{s} 
                  e^{-i\frac{2\pi}{2s+1}\beta\alpha} 
                  \psi(\alpha),\ \ (\beta=-s,-s+1,\ldots,s). 
              \label{eq_fourie}
\end{eqnarray}
Thus it is clear that the Fano operator $\Delta^{\rm S}(\alpha,\beta)$
in the spin labeling is related to our $\Delta(a,b)$ by 
the unitary transformation as 
\begin{eqnarray}
   \Delta^{\rm S}_{\alpha_1,\alpha_2}(\alpha, \beta) = 
   e^{-i\frac{2\pi}{N}(\alpha_1+s)s}
   \Delta_{\alpha_1+s,\alpha_2+s}(\alpha+s, \beta+s)
   e^{i\frac{2\pi}{N}(\alpha_2+s)s}, \nonumber \\
    ( \alpha,\beta,\alpha_1,\alpha_2 = -s,-s+1,\ldots ,s). 
\end{eqnarray}
In fact we can explicitly verify that if $\Delta(a,b)$ satisfies 
the four conditions, $\Delta^{\rm S}(\alpha,\beta)$ is also 
a solution to the 
four conditions in the spin labeling version, and vice versa.

However, it should be noted that the Fourier transformation
given in Eqs.(\ref{eq_fourie}),
especially in the spin labeling, 
is not unique. An interesting point in the spin labeling is
that the Fourier transform Eqs.(\ref{eq_fourie}) 
and its inverse transformation 
naturally lead to the anti-periodic
boundary conditions for a half-integer spin:
  $  \psi(\alpha+2s+1) = - \psi(\alpha) $ and  
  $  \tilde\psi(\beta+2s+1) = - \tilde\psi(\beta) $.
This reminds us of some flavor of the spin statistics theorem
as mentioned in \cite{cohendet}. Discussions 
on this issue, in conjunction with the reduction method 
introduced in Sec.~2, will be presented elsewhere \cite{hashimoto}.

\end{document}